\documentclass[11pt]{article}
\pdfoutput=1 
\usepackage{subfigure}
\usepackage[utf8]{inputenc} 
\usepackage[english]{babel}

\usepackage{graphicx}
\usepackage{amsmath}
\usepackage{authblk}
\usepackage[square,numbers]{natbib}

\title{Pre-stack deconvolution revisited}

\author{Jagmeet Singh}
\affil{Regional Computer Centre, ONGC Ltd, Vadodara,India}
\begin{document}
\title{Prestack deconvolution revisited}
\maketitle

\begin{abstract}
   While in the case of zero offset data with horizontal beds,it is clear that the recorded trace is a convolution of the wavelet and reflectivity series,
  the situation in the case of  nonzero offset is complicated by the fact that the direction of the incident ray is not vertical, the natural direction in which the reflectivity
  series is measured.Apart from a  simple and known consequence  that the periodicity of multiples, and thereby the validity of predictive deconvolution is lost at nonzero offsets,an important
 consequence that has been ignored is that very idea of the trace as a convolution of the wavelet and reflectivity gets 'limited' at nonzero offset.Moreover, the terms participating in this limited convolution or 'amalgam of amplitudes'
 arise from  wavelets along different rays getting reflected at different reflectors, resulting in a different `degree of convolution' across a gather and down a trace.We also examine tau-p deconvolution critically, which relies on  the assumption that the slant stack represents a convolution of zero offset reflectivity
 at a particular p and the wavelet.We find this assumption to be wrong. 
\end{abstract}

\section{Introduction}

While in the case of zero offset data with horizontal beds, it is clear that 
the recorded trace is a convolution, say $w_{1}$ $R_{3}$ + $w_{2}$ $R_{2}$ +
$w_{3}$ $R_{1}$, where w and R with their indices represent the wavelet and reflectivity series at different times \cite{Yilmaz.seismic.2001}, the situation in the case of finite offset is not so simple. For finite offset, the direction of 
the incident ray is not vertical, the direction along which one normally measures the reflectivity
series.A question that arises naturally is:- 'what is the direction along which time/reflectivity series is measured-is it the slant direction along 
the incident ray or the vertical direction?'.We examine this question as well the effect of the two directions involved on the convolution itself.Origin of time clearly has to be the moment of initiation of blast.

In the section below namely \textit{Different Ray Paths} , we  examine the situation from two different standpoints \textit{i})a
single ray emanating from a source hitting different reflectors at different CMP's and \textit{ii}) multiple rays maintaining the same CMP. We restrict ourselves to the case of horizontal reflectors in this paper.We shall examine whether the idea of convolution survives or not--whether we have to think in terms of convolution or consider the
recorded amplitude at a particular time as just an 'amalgam' of amplitudes reflected from nearby reflectors.We need to introduce the idea of  'degree of convolution', which varies from trace to trace(i.e. with offset) and with time in a particular trace.This leads to the conclusion  that a single deconvolution operator can not deconvolve an entire trace, and also that gather level deconvolution(e.g. shot deconvolution) is not meaningful.In the next section,we take a critical look at nonzero offset predictive deconvolution and highlight its known limitations. A section looking critically at the idea of  $\tau$-$p$ deconvolution then follows. Deconvolution in this domain is based on the premise that a p trace is convolution of the wavelet 
and reflectivity in zero offset time at the particular p. We find this assumption to be wrong.Finally, we draw the
main ideas of the paper in the last  section \textit{Conclusions}.

\section{Different ray paths}

\par{ As discussed above, a slant incident ray meets the  series of reflection coefficients in the slant direction, whereas the series itself is defined in the vertical direction.
Figure~\ref{fig:fig1} shows a single ray path that meets different reflection coefficients $R_{1}$,$R_{2}$ and $R_{3}$ at different times along the slant path.When we write h(t) = w(t) * R(t),what  is the origin of
time the for the three quantities involved? Obviously,all the three quantities
must refer to the same origin of time ---the moment of  origin of  blast. Even the reflectivity series
has to be measured/sampled along the slant direction i.e. the ray direction. If
one  assumes that  the reflectivity series is the series of reflection
coefficients measured in the vertical direction, the convolution does not stand. In
figure 1 , the difference in times between w$_{1}$ and w$_{2}$(in the convolution sum $w_{1}$ $R_{3}$ + $w_{2}$ $R_{2}$ +
$w_{3}$ $R_{1}$) must match the difference in times between R$_{3}$  and R$_{2}$ which happens when time axis
for w and R is the same. If we insist on defining reflectivity in the vertical direction independent of the direction of the
incident ray,the resulting 'amalgam of amplitudes' is something like
\begin{equation}\label{1}
h(t)=\int_{-\infty}^{\infty} w(\tau) R((t-\tau)\sin\theta)d\tau
\end{equation}

where $\theta$ is the angle between the incident ray and the horizontal.This is clearly not a convolution-- only when $\theta$ is 90$^{\circ}$ i.e. the case of normal incidence,  convolution is restored.A simple solution then is to redefine reflectivity series as the one sampled along the slant direction of the incident ray itself rather than the vertical direction.However, as shown in Figure 1 below, we find that the
reflections ( $w_{1}R_{3}$, $w_{2}R_{2}$, $w_{3}R_{1}$)that would normally be 
received at a single receiver (in the case of zero offset), are received at
different receivers. So, the reflected amplitudes that were supposed to exist in the convolution sum at a single location are in fact being received at different receivers.
A possible solution to restore the convolution could be to simply
add across the traces at R1,R2 and R3 i.e. a kind of local stack.But why generate a convolution when it's
not there in the first place.
\begin{figure} 
\centering
\includegraphics[height=20 cm,width=\textwidth]{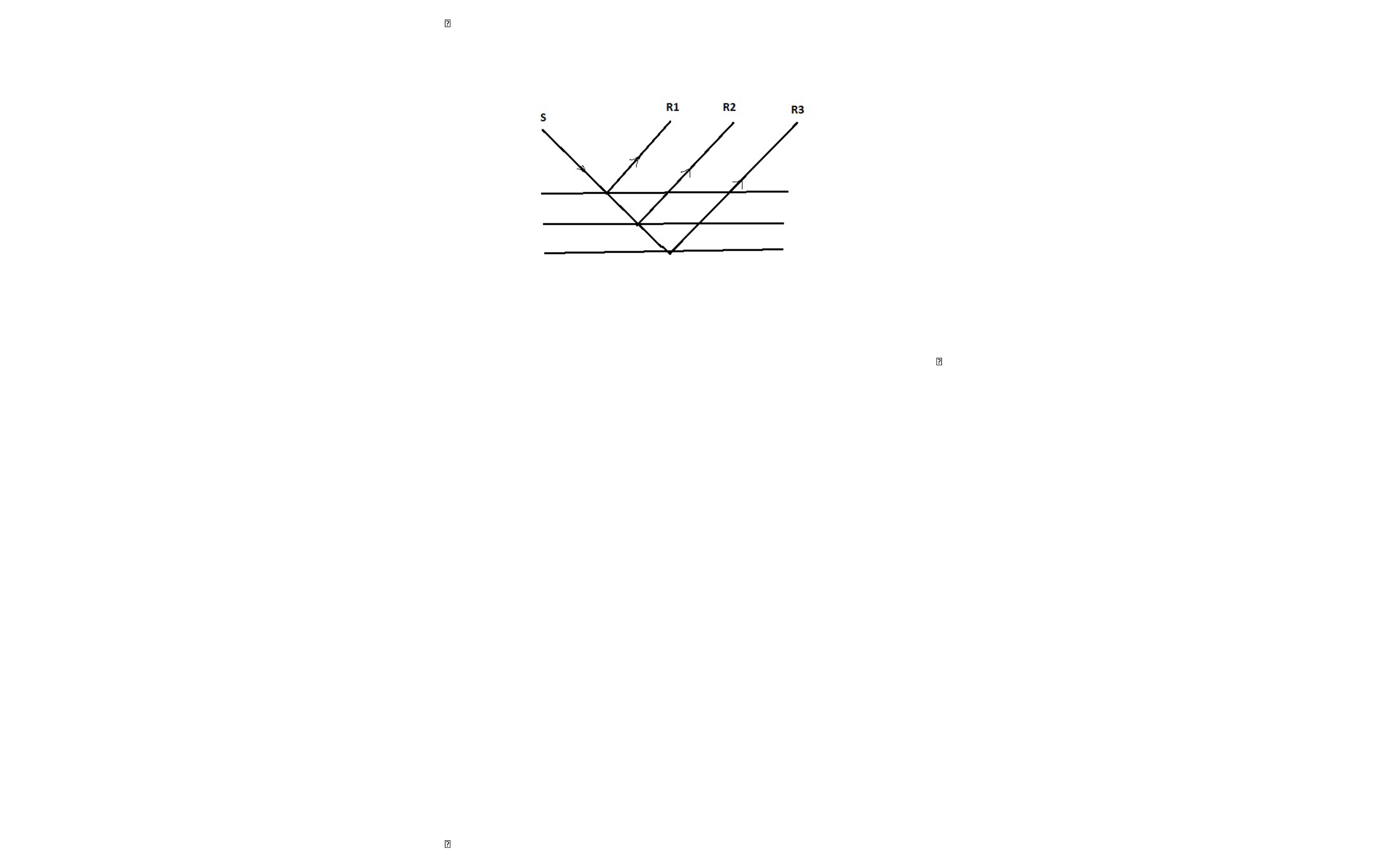} 
\caption{A single ray path getting reflected and received at multiple receivers} 
\label{fig:fig1}
\end{figure}
}

\par
\begin{figure}
\centering
\includegraphics[height=20 cm,width= 20 cm]{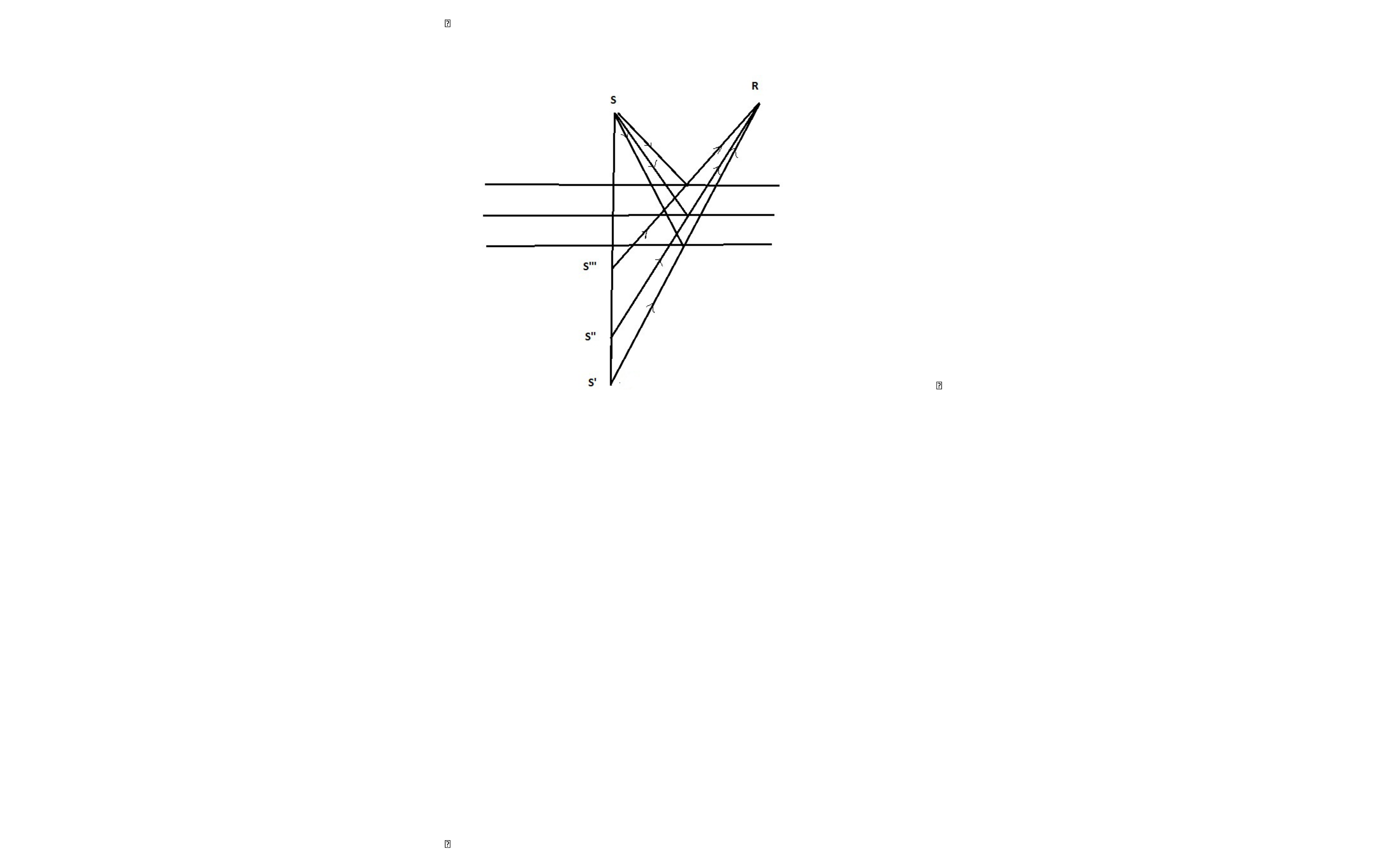}
\caption{Multiple ray paths getting reflected and received at a single receiver}
\label{fig:fig2} 
\end{figure}

{

 Let us now look at the picture from another angle. Instead of looking at a
single incident ray, let us examine what is received at a single receiver from
a source (Figure~\ref{fig:fig2}). The source S is reflected at the three reflectors shown to  
produce virtual sources S''',S' and S'. Three ray-paths result corresponding to
different depths but the same CMP. Can such a trace  be looked upon as a
convolution between the wavelet  and the reflection coefficient series?  The
important difference here is that there are multiple ray paths  involved. In  the
case discussed above, reflectivity series was measured along a single ray, whereas, in this case, it is being measured
along different rays.It is indeed a problem that there is no single direction along which reflectivity
is measured.However,whatever may be the awkward direction-changing sequence of  reflection coefficients,we may still call it as a 'convolution'(limited convolution) between the wavelet and the reflection field(field of reflection coefficients) as measured from the source point.Since time differences between two oblique rays in Figure 2(different from the time differences between corresponding vertical rays) must match time differences between the wavelets at the two locations,
there will be a diiferent set of terms in the convolution sum.

We shall see below that we need to introduce the idea of 'degree of convolution' which in the slant case is different from that in the vertical case.For a single trace recorded at a receiver, the angle changes with depth and the terms participating in the convolution also change i.e. shallow reflectors may be more convolved than the  deeper ones.As we go deeper and deeper, we go closer to the vertical case. 
}

 To quantify the two dimensional nature of the problem(case of slant rays), we write the argument of the wavelet as
$(t- r/c)$  (instead of $t$)  where $$r=|{\vec r}|$$  is the
distance traveled from the source point to the reflector and c is the wave
speed. We define a hypothetical trace h(t), summing over all amplitudes received at the surface from different reflection points $\vec r$ at two way time t  as 
\begin{equation}\label{2}
h(t)= \int{}w(t-r/c) R(\vec r/c)d(\vec r/c),
\end{equation}
where we have used argument of reflection coefficient as $\frac{\vec r}{c}$ instead of $\vec r$ to let arguments of both w and R be in time.By 
$d(\vec r/c)$, we mean $rdr d\theta/c^2$ and $R(\vec r/c)$ means $R(x/c,z/c)$.

Why we call it hypotheical is because it is never recorded, as it  sums over all possible positions $\vec r$($r<=ct$).If the x component of $\vec r$ is not fixed, the reflections are received at different receivers.
So, in the above equation we need to integrate over z keeping x fixed if we need to find the amplitude received at one receiver i.e.
\begin{equation}\label{eq:3}
h(t,x)=\int{}w(t-\frac{\sqrt{(x^2+z^2)}}{c}) R(\frac{x}{c},\frac{z}{c})dz/c.
\end {equation}
For the flat reflector case that we are considering, dependence of R on x in the above equation may be dropped.Clearly equation \eqref{eq:3} does not represent a convolution.So, what's wrong?
The answer is nothing--the recorded trace is not a convolution between the reflectivity \textit{measured in the vertical direction} and the wavelet! There is, however, another way of looking at the problem --the reflector is met when t=r/c  and the reflectivity is  referred to time t with the origin of time as the moment of initiation of blast.With reflectivity series defined in two way time r/c, equation(\eqref{2}), for the case of constant x(single receiver) may be re-written as 

\begin{equation}\label{4}
h(t,x)=\int{}w(t-\frac{\sqrt{(x^2+z^2)}}{c}) R(\frac{\sqrt{(x^2+z^2)}}{c},x)\frac{z}{c\sqrt{(x^2+z^2)}}dz,
\end {equation}

where we have written arguments of R as the two way time $\sqrt(x^2+z^2)/c$ as well as position $x$.We need an additional qualifier x  because same time($\sqrt(x^2+z^2)/c)$ may correspond to two different x(or z).Further,we have used 
$$d(r/c)=\frac{zdz}{c\sqrt{(x^2+z^2)}},$$ valid for the case of constant x.Defining $$\tau=\frac{\sqrt{(x^2+z^2)}}{c},$$ we see that the above may be re-written as:
\begin{equation}\label{5}
h(t,x)=\int{}w(t-\tau) R(\tau,x)d\tau,
\end {equation}

which looks like a convolution at least for a constant x.The reflectivity series is sampled differently at different x, so that we need an additional qualifier x in R,
even for the case of horizontal reflectors.A question that naturally arises is 'which equation ( \eqref{eq:3} or \eqref{4}) describes correctly the amplitude recorded at x?'.Both equations \eqref{eq:3} and \eqref{4} need a slight modification--both have dimensions of time and need to be normalized.We need to divide \eqref{eq:3} by two way vertical time $t_0$ corresponding to time $t=R/c$ (where R is the distance up to reflector met by wavelet amplitude $w_0$) i.e. $t=Z/c$,  and \eqref{4} by two way time t.When we do this, it is easy to see that equations \eqref{eq:3} and \eqref{4}are essentially the same--they are two ways of looking at the same problem.Although, for a constant x(and flat reflectors), the recorded trace given by \eqref{4} is mathematically  a convolution of the wavelet and \textit{reflectivity measured in two way time}, we need to understand the limitations of such a definition.There is no unique direction along which the reflectivity (or the field of reflection coefficients) is measured, resulting in dense or sparser sampling at different depths i.e. the degree of convolution varies with depth in a trace.

If we look back at the relation $$d(r/c)=\frac{zdz}{c\sqrt{(x^2+z^2)}}$$ or $$d(r/c)=\frac{dz}{c\sqrt{(1+x^2/z^2)}}$$, we see that as we go deeper i.e. z increases, we get closer to the vertical difference dz/c and at shallower z, the difference is lower than $dz/c$ meaning thereby more terms in the convolution, or greater degree of convolution.That is, shallower reflectors are more convolved than the deeper ones in a nonzero offset trace.Though with equation \eqref{5}, we have managed  to give our amalgam of amplitudes, mathematically,  the shape of a convolution--we have to understand the limitations of such a definition.With time varying degree of convolution in a trace represented by equations \eqref{4},\eqref{5}, we can not expect a single deconvolution operator to deconvolve the whole trace, and with different reflectivity series across  a gather(although R is not a function of x), a single deconvolution operator for the whole gather would be problematic. For example,in the case of surface consistent deconvolution,we expect a single deconvolution operator for a shot gather(other things being equal i.e. identical receivers in identical surface conditions)--this is clearly not right. For single trace deconvolution, another way of looking at the problem is that the whiteness of the reflectivity series is  disturbed by the irregular sampling of the reflection series by slant rays--making it difficult to deconvolve the trace.  

 One may be tempted to think that NMO which maps
two way slant time t  onto two way vertical time t$_{0}$ would solve the
problem i.e. restore the 1D convolution suggesting NMO befere deconvolution. But the constituents of the amalgam remain the same--it's just that the amalgam  $h(t)$ is mapped onto $h(t_{0})$
This will not be the same as 1D convolution recorded in the case of zero offset.

\par{
What we arrived at intuitively in the two cases described above(Figures 1 and 2) is  well encapsulated in the equation above.The equation \eqref{eq:3} corresponds perfectly to the case 2 (Figure 2) discussed above.To cover case 1, we need to fix the angle $\theta$ 
in equation\eqref{1} so that we get
\begin{equation}\label{6}
h(t,\theta)= \int{}w(t-r/c) R(r, \theta)d( r/c),
\end{equation}  where we have dropped the integration over $\theta$.Note that this sum of amplitudes is also not recorded as its constituents are received on different receivers.
We also need to consider the Fresnel zone at the surface in case 1.If the constituents of the amalgam of amplitudes reaching different surface locations fall within the Fresnel zone, we may as well take them to be recorded by a single receiver.
 However, for the sake of simplicity, we shall assume, here, the Fresnel zone to be small enough, and receivers well separated so as not to consider this case.
}

\section {Predictive deconvolution}
\par
\begin{figure}
\centering
\includegraphics[height=20 cm,width=20 cm]{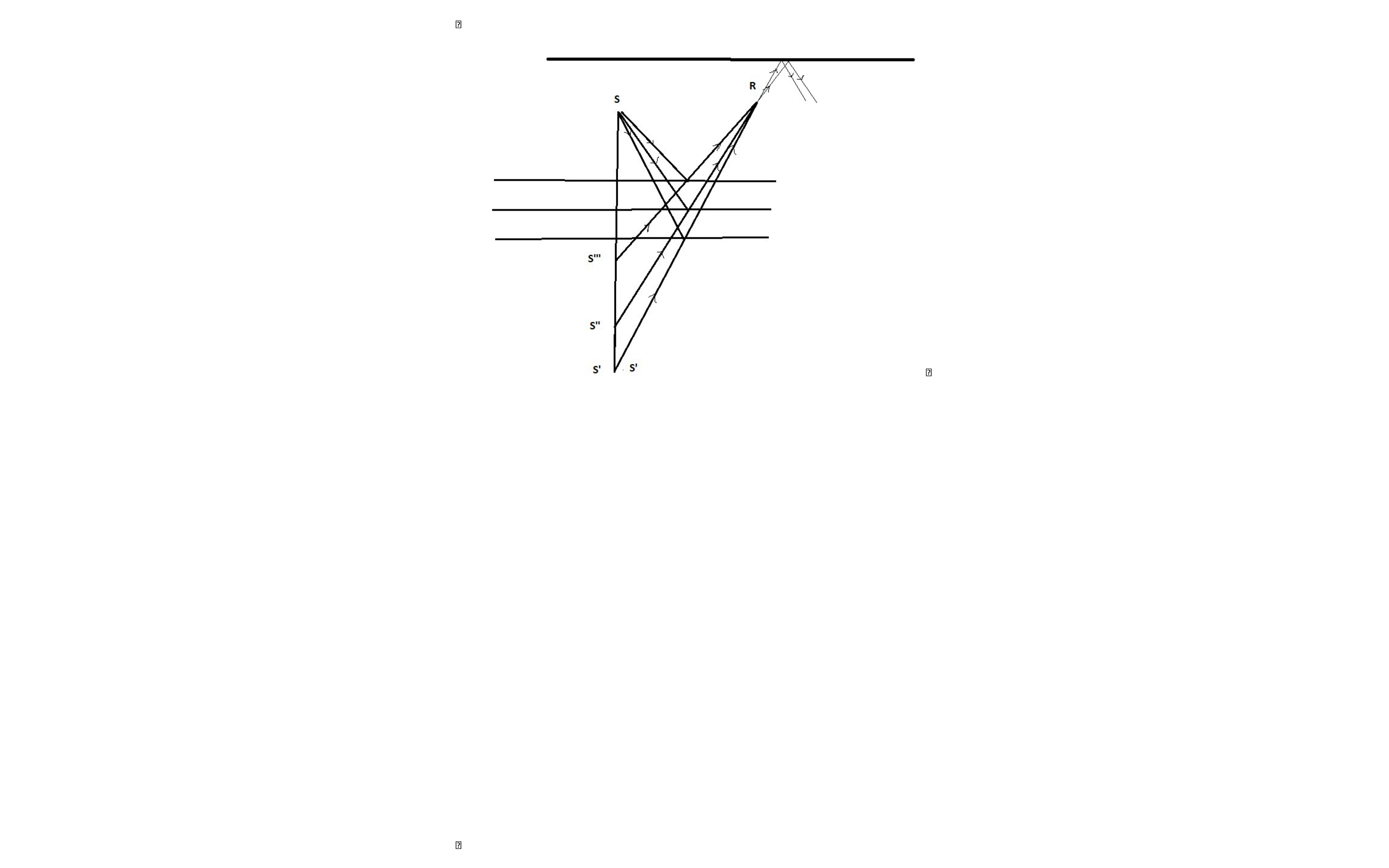}
\caption{Multiple generation from sea surface}
\label{fig:fig3}
\end{figure}

{
Although the limitations of predictive deconvolution in t-x domain for multiple attenuation at non-zero offsets are well known, for the sake of completeness,we  delve a little into the subject, before we move to the $\tau-p$ domain.
All predictive deconvolution algorithms available in the seismic industry assume the simple 1D convolution $h(t)=w(t)*R(t)$, and also  that there is a predictable part in this convolution(the multiples) which can be subtracted. Now that we agree that what is recorded at a single receiver R is 
some amalgam of amplitudes arising from reflections from different rays or a convolution in limited sense, the
next questdon that arises is `can  we retain predictive
deconvolution or a similar idea for multiple attenuation?'. Obviously, for predictive deconvolution to work, we
need predictability apart from a convolution.We
examine predictability here \& we shall restrict ourselves to the case of
horizontal reflectors. Predictability of the wavelet generally comes from a
source of reverberation like the sea surface. Imagine sea surface lying just
above source S and receiver R shown in Figure~\ref{fig:fig3}  and the first reflector being the
water bottom.We see that the periods of the multiple energy corresponding to different rays shown are different and also
received at different  receivers, so we focus instead on multiple energy received at a particular receiver.We see that multiples received at a particular receiver would be generated from different CDP's(whereas primary energy is from
 the same CMP (for the case of horizontal reflectors)  and the periods of multiple from different reflectors would be different.
}

\begin{figure}
\centering
\includegraphics[height=20 cm,width=16 cm]{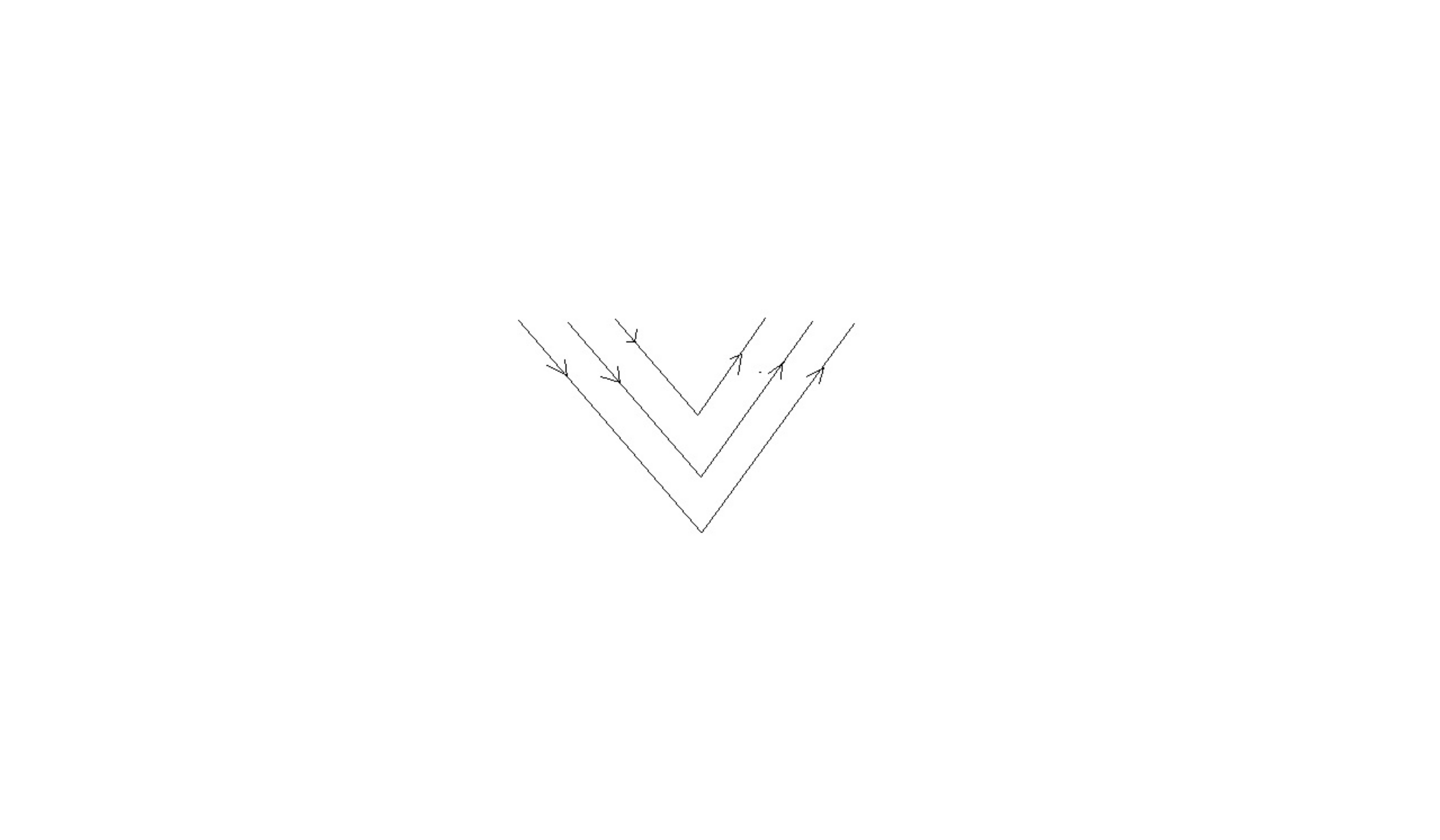}
\caption{Rays with same angle}
\label{fig:fig4}
\end{figure}

{\raggedright
Cleadly, predictive deconvolution in the pre-stack case as a multiple attenuator
fails.  Even if we consider a single trace as a convolution of the wavelet and
reflectivity series , the period of multiples is not fixed and
predictive deconvolutnon as a multiple attenuator will not work. So, what is the way out?In Figure~\ref{fig:fig4} , we show rays with same angle recorded at different receivers at different times.Such energy would be difficult to isolate,
 but would be convolutional.However, multiple energy corresponding to these primaries would be received at different receivers--so we have to reject this idea as well.
}

{\raggedright
We mentioned possible summation across traces while discussing the case of single ray---this reminds one of
the slant stack[1]. A plane wave is received by multiple receivers--- is the stack
along the plane ( of the plane wave)  a convolution? A beauty about the plane
wave is that the period of the mutiple is fixed for a particularly directed
plane wave i.e. a plane wave with a particular p value. So, in this case, does
predictive deconvolution work?We examine this question in the next section.
}

{\raggedright
\textbf{Slant Stack Relooked}
}

{\raggedright
An array of sources with suitable delays is expected to generate a plane wave.
Since we have  a single source in a shot gather, we simulate a plane wave by
introducing the delays in the array of receivers instead. What is however recorded by
the receivers is not pure plane waves as they have been altered by the reflectors.A 
more rigorous interpretation of the case in hand is that of a point source(time harmonic) which is  a summation of plane waves via the Sommerfeld-Weyl
integral \cite{GEO49-09-14951505}.A real wavelet of finite duration is, of course, however simulated by a superposition of such time harmonic point sources.We assume that  the modification of these plane waves by the reflectors
 is  such that the sum over delayed receivers at a given $p$ yields a convolution sum of these plane waves with the reflectors or in other words a convolution of the wavelet and the reflectors in zero offset time,
so that a subsequent deconvolution in the $\tau-p$ domain would be justified.However, as the angle of incidence increases, the reflected plane waves are recorded by less and less of the spread 
of receivers--a limitation that we need to keep in mind.Moreover, even in the case of infinite spread, we need to see mathematically if the slant stack is really a convolution of the wavelet and reflectivity in zero offset time or not.
we work it out below.Starting with the definition of trace at a particular x  
\begin{equation}\label{7}
h(t,x)=\int{}W(t-t') R(t',x)dt',
\end {equation}
where $t'=\frac{\sqrt(x^2+z^2}{c}$.Making the substitution $t=\tau + px$, and summing over x, we get
\begin{equation}\label{8}
H(p,\tau)=\int{}W(\tau+px-t') R(t',x)dt'dx,
\end {equation}
Introducing 
\begin{equation}\label{9}
W(t)=\int W(\omega)\exp(\iota\omega\tau)d\tau,
\end{equation}
 we write equation \eqref{8} as 
\begin{equation}\label{10}
H(p,\tau)=\int W(\omega)\exp(\iota\omega\tau)\exp(\iota\omega(px-t'))R(t',x)d\omega dx dt',
\end{equation}
we get 
\begin{equation}\label{11}
H(p,\tau)=\int W(\omega)\exp(\iota\omega(\tau+px)R(\omega,x)dx d\omega,
\end{equation}
where we have used the identity
\begin{equation}\label{12}
R(\omega)=\int R(t')\exp(-\iota\omega t')dt'
\end{equation}
Note that factors of $2 \pi$ etc. have been left out as these would cancel out in the use of inverse and forward transforms given by equations \eqref {9} and \eqref{12}.Equation \eqref{11} wonderfully describes $\tau-p$ transform as a sum of plane waves with amplitudes as a product of $R(\omega,x)$ and $W(\omega)$--this is meaningful only when dependence of R on x can be dropped.With a little re-arrangement, we can show that  
\begin{equation}
H(\omega,p)=\int W(\omega) R(\omega,x)\exp(\iota kx)dx
\end{equation}
 or
\begin{equation}\label{14}
H(\omega,p)=W(\omega) R(\omega,k)=W(\omega) R(\omega,p \omega).
\end{equation}
The above equation (\eqref{14}) falls short of convolution because of different arguments ($p$ and $k$)on the left and right hand sides.This shows that the slant stack is not a convolution of the wavelet and reflectivity in zero offset time as is commonly believed, rendering the idea of $\tau-p$ deconvolution to recover the reflectivity series to be  incorrect.But  at least, the second condition (discussed above) for predictive deconvolution namely periodicity of multiples for a particular p is met,so $\tau-p$ predictive deconvolution for multiple attenuation,though not strictly correct, is  more meaningful than predictive deconvolution in $x-t$ domain.      

}

\section{Conclusions}
\par{Although there is nothing wrong with deconvolution of a zero offset trace,one needs to be very careful while extending the idea to the case of nonzero offset, where the reflection coefficient series becomes a function of both time and offset, even for the case of horizontal reflectors.Gather level deconvolution i.e. a single deconvolution operator for a gather is not meaningful, and even for a single trace, a single deconvolution operator can not deconvolve an entire trace .Predictive deconvolution in t-x domain as a multiple attenuator is also not meaningful.We have also examined the slant stack critically, and found that deconvolution in this domain as a tool to recover the reflection coefficient series R(t) is not meaningful.But since multiples are periodic in this domain, predictive deconvolution may still be tried as a statistical tool for multiple attenuation.


\medskip



\end{document}